# Engineering an Interstellar Communications Network by Deploying Relay Probes

**JOHN GERTZ**[1] **& GEOFF MARCY**[2] [1]Zorro Productions, 2249 Fifth Street, Berkeley, CA, 94710; [2]Space Laser Awareness, 3388 Petaluma Hill Rd, Santa Rosa, CA 95404

**Email** jgertz@zorro.com

We develop a model for an interstellar communication network that is composed of relay nodes that transmit diffraction-limited beams of photons. We provide a multi-dimensional rationale for such a network of communication in lieu of interstellar beacons. We derive a theoretical expression for the bit rate of communication based on fundamental physics, constrained by the energy available for photons and the diffraction of the beam that dilutes the information by the inverse square law. We find that meter-scale probes are severely limited in their bit rate, under 1 Gbps, over distances of a light year. However, that bit rate is proportional to the 4th power of the size of the optics that transmit and receive the photons, and inversely proportional to the square of the distance between them, thus favoring large optics and short separations between nodes. The optimized architecture of interstellar communication consists of a network of nodes separated by sub-light-year distances and strung out between neighboring stars.

## 1 INTRODUCTION

Bracewell postulated in 1960 that extraterrestrials with technology (hereafter "ET") might communicate over interstellar distances by using physical machines (hereafter, "probes") [1,2], opening discussion of their reproduction and communication protocols [3,4]. Rose & Wright [5] calculated that information might be encased within a probe in the form of inscribed matter, e.g. bits stored on the atomic scale, at a lower energy cost than were the same information transmitted by means of radio waves or lasers from one star system to another. Gertz [6] organized a list of reasons for ET to prefer interstellar probes over the use of electromagnetic waves transmitted from its home star system, among which were:

- An ET probe with AI in our Solar System might surveil Earth's omnidirectional electromagnetic transmissions, learning our languages, science, and cultures.
- A local ET probe might enter into a dialogue with Earth in near real time, transmitting in human languages to improve mutual comprehension.
- The probe might communicate its findings back to its progenitor civilization.
- Alternatively, a fully autonomous local ET probe might outlive its progenitor civilization, obviating the need for coexistence of the two civilizations, as represented by Frank Drake's *L* factor.
- Communicating with Earth only by means of a local probe avoids revealing the spatial coordinates of its progenitor civilization, thus eliminating the inherent security danger of such information.
- The AI probe might assess any potential danger posed by life on the target planet, providing its progenitor civilization with interstellar situational awareness.

If an ET probe located within our Solar System is currently surveilling Earth's artificial EM leakage, it might record internet transmissions as well as television, military and other human communication. It might synthesize, analyze and reduce this vast data trove before transmitting its work product back to its progenitor civilization many light years (hereafter sometimes, "ly") away. The amount of analyzed data the probe might transmit would likely value a high bit rate to transmit a meaningful amount of information within a reasonable amount of time. The purpose of this paper is to examine the engineering challenges facing ET probes in constructing effective communication links.

## 2 ET'S POSSIBLE SOLUTIONS TO TRANSMISSION CHALLENGES

In this section we examine four hypothetical solutions available to the local ET probe to the problem of communicating information back to its progenitor civilization and/or to a network of linked ET civilizations.

### 2.1 Unassisted Star-to-Star Communication: Constraints from Physics

Gillon [7] and Gertz [8, 9, 10, 11] postulated that communications among civilizations might be facilitated by a class of probes that act as communications nodes. These might be placed in orbit around some, or even most stars.

We consider here a local ET probe of a certain size sending information to a probe at another star over distances measured in light years. Each probe has a certain energy budget for communication, which is spent by sending photons. Physics imposes two fundamental constraints on the amount of information that can be transferred from one probe to the other.

First, the energy budget limits the number of photons that can be sent, as each photon carries away an energy, hv. Here h is Planck's constant and ν is the frequency of the light. Second, the size of the probe dictates the ultimate narrowness of the beam of photons, set by diffraction, that concentrates the photons on their journey toward the receiver. The total amount of energy available to a local ET probe, along with the narrowness of the beam, will determine the total amount of information that it can send to a receiver, including to other probes within the galactic internet or to its progenitor civilization.

### 2.1.1 Probe-to-Probe Communication: Solar System to Alpha Centauri

To set the scale of parameters, we consider a benchmark case consisting of an ideal light beam launcher having a diameter of 1 meter located in our Solar System and aimed at Alpha Centauri 4.3 ly away. Diffraction of the light as it exits the launcher will spread the beam into an ever-widening cone. For optical wavelengths (400-700 nm), such a diffraction-limited, light beam would have an opening angle of ~0.07 arcseconds.

The conical, expanding beam upon arrival at Alpha Centauri will have a footprint diameter of 0.1 AU (~15 million km). That diameter will be fatter for wavelengths longer than optical light, and narrower for UV or x-ray light, due to diffraction of light at the aperture of the beam launcher. The key physics reality is that optical photons emitted by this ideal probe will be spread over a circular footprint 15 million kilometers in diameter, greatly diluting their intensity. The area of that circular footprint at Alpha Centauri is $2 \times 10^{20}$ square meters. Thus, a receiver at Alpha Centauri having a collecting area of 1 square meter would acquire only 1 photon for each $2 \times 10^{20}$ photons emitted by the probe in our Solar System – phenomenal inefficiency.

A finite energy budget would result in a maximum number of photons that can be emitted (of some certain energy), hence limiting the amount of information that can be sent. As a benchmark, we consider 100 kilograms of Uranium-235 that yields 8300 TeraJoules of energy (= 2.3 million Megawatt-hours). A typical ultraviolet photon has an energy of ~10 electron volts (= $1.6 \times 10^{-18}$ Joules). Thus, 100 kg of U-235 yields $5 \times 10^{33}$ photons.

One photon yields only approximately one bit of information. The actual yield might be less given additional requirements for error correction and redundancy, or it might be more if information is conveyed not simply by means of photon presence (1) or absence (0) but additionally by the exact timing between pulses, the exact frequency of each photon, or its polarization. For current purposes, we will consider the effects to average out such that one photon = 1 bit. At first glance, $5 \times 10^{33}$ bits of information seems like a lot. However, the previous paragraphs show the benchmark case results in only 1 photon out of $2 \times 10^{20}$ photons will be captured by a 1-meter receiver. Immediately we see that of the emitted $5 \times 10^{33}$ photons, only ~$2 \times 10^{13}$ photons will be received by a 1-meter class receiver. The total information received at Alpha Centauri is roughly 20 terrabytes – the amount of storage on a typical external hard drive for a laptop computer. A finite energy budget and diffraction of the beam over interstellar distances limits the total information transfer to humbling home-computer levels.

### 2.1.2 Transmitter-to-Receiver Communication: The General Case

The previous benchmark example shows that a challenge facing a meter-scale ET probe is that for a given amount of energy available to it, the bit rate constraint may be prohibitively low to allow for voluminous star-to-star communication. Nearly all of the energy will miss the receiver in an enormously wide light cone. Diffraction is the enemy of interstellar communication, but it offers astronomers the hope of serendipitously detecting the photon spillover of ET communications not meant for Earth.

This tension between energy budget and information transfer holds even if the energy is not expended at a steady rate but is husbanded to be released mostly after the probe has detected artificial electromagnetic radiation from the planet it is surveilling and only then sends the preponderance of its inventory of information. The diffraction of light, along with interstellar distances conspire to waste nearly all the energy budget devoted to communication.

Here we derive a fundamental expression for the amount of information that can be sent from one transmitter to a receiver some distance away. We consider the energy required to create individual photons and the cone-shaped beam caused by diffraction as the photons leave the beam-launch aperture of the emitter.

The diffraction-limited beam opening angle, θ, of the light cone depends inversely as the diameter, D, of the beam where it is launched and is proportional to the wavelength, λ of the light. The full, diffraction-limited opening angle is given by,

$$\theta = 2\lambda/\pi D \qquad (1)$$

The beam has fuzzy edges, so the factors of 2 and π are not important. As a useful reference, a diffraction-limited beam launcher of diameter 1 meter (a "beam waist" of 1 meter), emitting visible ("optical") wavelengths (500 nm), will emit a cone of light with an opening angle of 0.066 arcsec. Launched from the Solar System, the beam makes a footprint at Alpha Centauri that is 0.089 AU across.

One may easily derive the general expression for the average flux of photons, F, having wavelength, λ, launched from a diffraction-limited emitter of diameter, D, and received a distance, d, away. The geometry of the launcher, the beam divergence, and the receiver are shown in Figure 1.

The functions of the ET probe will be constrained by the total amount of energy it brought along as payload. The energy may be in various forms such as fissionable nuclear fuel (U-235 or Pu-244), fusion-ready hydrogen, deuterium or tritium, or some matter/antimatter material in storage tanks. We consider the possibility of additional energy acquired later below.

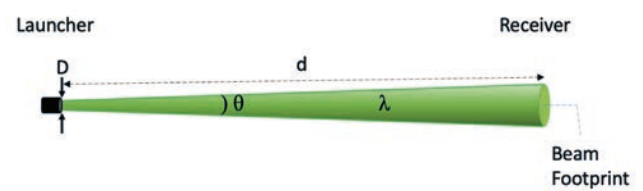

**Fig.1** The geometry of the beam launcher (at left) with an aperture of diameter, *D*. Diffraction causes the light of wavelength, *λ*, to spread out with opening angle, *θ*. After traveling a distance, *d*, the beam makes circular footprint area, *1/π (λd/D)²*, presumably much larger than the receiver.

The total initial energy stored as payload, $E_i$, may be used to generate photons (and perform other tasks such as station keeping). This energy $E_i$ is expended during a transmission lifetime, $L$.

We compute the photon flux as follows. The maximum number of photons that can be emitted is simply the total energy available, $E_i$, divided by the energy per photon, $hc/\lambda$. Those photons are eventually spread over the area of the circular beam footprint that results from the diffraction-limited beam opening angle (Eq.1) and the distance to the receiver, d, giving a circular footprint area, $1/\pi\,(\lambda d/D)^2$. The ratio of the total number of photons to that footprint area gives the photons per unit area at the location of the receiver. Dividing that ratio by the time spent emitting, $L$, gives the average number of photons per unit area per unit time, i.e., the photon flux (in units of photons m$^{-2}$ s$^{-1}$). Assembling the components above, the average photon flux during a time $L$ is:

$$F = \pi/hc\, E_i\, D^2 / (L\, \lambda\, d^2) \qquad (2)$$

Equation 2 is a general expression for the flux of photons, measured in photons per square meter per second, for an emitting probe having aperture diameter, $D$, and energy budget, $E_i$, emitting photons of wavelength, $\lambda$, over a distance, $d$, during a transmitting lifetime, $L$.

Note that the *photon* Flux at the receiver has the following dependencies:
- Photon Flux decreases with the square of distance, $d^2$, as usual. Smaller separations increase the photon flux.
- Photon Flux increases linearly with $E_i$, the energy available to generate photons.
- Photon Flux increases with $D^2$ because of the smaller ("tighter") diffraction-limited opening angle, $\theta$, of the beam. The flux increases as the square of $D$ because small beam size creates a small 2-dimensional footprint area.
- Flux decreases with $L$, the lifetime of the probe's continuous transmissions. A longer lifespan spreads the available photons over a longer time, decreasing flux.
- Flux decreases with $\lambda$ *(lambda)*. Longer wavelengths cause greater diffraction-limited beam opening angle, diluting the beam flux at the receiver.

Photon flux benefits from higher energy budget $E_i$, larger emitting aperture $D$, shorter lifespan $L$, and shorter transmission distance, $d$. Importantly, the photon flux is higher for shorter wavelength such as UV or X-rays, due to the tighter diffraction-limited beam, both carrying implications for optimization of information transfer rates.

We include now the collecting aperture of the "telescope" at the receiver. For probes of a given size, the diameter of the transmitter aperture, $D$, and that of the receiver aperture, are likely to be roughly similar in size. It makes little sense for one to be several orders of magnitude larger than the other. Indeed, the beam launcher and beam receiver may be interchangeable, switchable if sending or receiving.

Therefore, we adopt a receiver collecting aperture that has an circular area, $\pi\,(D/2)^2$. We multiply the photon flux in equation 2 by this receiver aperture area to derive the number of photons collected per second:

$$\textit{Photon Collection Rate} = \pi^2/4hc\, E_i\, D^4 / (L\, \lambda\, d^2) \qquad (3)$$

Equation 3 is a general expression for the optimal photon rate between a transmitter and receiver, with units of photons per sec. It involves the aperture diameters of the transmitter and receiver, $D$, the distance between them, $d$, the energy budget $E_i$, the lifetime of transmission, $L$, and the wavelength, $\lambda$.

Equation 3 is a general probe-to-probe communication expression. It applies to any pair of probes for which the diffraction-limited beam overfills the receiver. Thus Equation 3 is a general expression for the information bit rate between any two identical probes.

Equation 3 shows that the information transfer rate increases with the 4th power of the size of the beam launcher and receiver, $D^4$. Thus, all design and engineering tradeoffs in the construction of optimal probes place a premium on transmitter and receiver size, despite the added mass and cost. Further, Equation 3 shows the information transfer rate is inversely proportional to the square of the distance, $d^2$, between the probes. Thus, tradeoffs in constructing the network of probes must favor shorter separations, despite the greater number of probes to fill the entire distance. Clearly, short wavelengths are favored, i.e. UV or x-ray. Thus, Equation 3 portends Galactic networks of mutually communicating probes that are large, closely spaced, and operating at short wavelengths.

### 2.1.3 Benchmark Properties of Probes for Communication

Equations 2 and 3 offer an opportunity to insert representative values for certain probes to construct benchmark cases. We consider a photon launcher with a diameter of 1 meter, emitting photons of wavelength 1 μ (near IR), with an initial energy budget of $9 \times 10^{17}$ Joules = $2.5 \times 10^{11}$ kWh. This is the energy obtained by complete conversion of a mass of 10kg to energy, i.e. $E_i$ = 10kg * $c^2$. We consider transmission over five distances, 0.01, 0.1, 1, 10, and 100 ly. Table 1 shows the resulting flux of photons at the receiver in units of photons per meter$^2$ per second.

Table 1 shows the resulting photon flux for a benchmark probe equipped with a 1-meter photon launcher, an initial photon energy budget of $2.5 \times 10^{11}$ kWh, and emitting photons at a wavelength of 1 μ. Table 1 lists the photon flux at the receiver for two different transmission lifetimes, $L$ = 100 yr and $10^6$ yr. For example, the benchmark 1-meter photon launcher equipped with a photon energy budget of $2.5 \times 10^{11}$ kWh can produce a photon flux at a receiver located 1 ly away of $5.1 \times 10^7$ photons per m$^2$ per sec during 100 years. If the same system expends its photons during a million years, the average flux is only 5,100 photons per m$^2$ per sec. Of course, over greater (or smaller) distances, the resulting flux decreases (increases) as the square of distance, as shown in Table 1.

### 2.1.4 Communication Bit Rates and Realistic Probes

Table 1 shows that two probes of 1-meter size can send and receive at most ~50 million optical photons/sec during a 100 year transmitting lifetime, given a $2.5 \times 10^{11}$ kWh energy budget (= 10 kg * $c^2$) when spaced by a mere 1 ly. This 50 million photons/sec rate is a stark constraint, defined by physics not engineer-ing. The communication bit rate is thus constrained.

**Table 1: Photon Flux for 1-meter Launcher for λ = 1μ and Energy = 2.5 x 10¹¹ kWh**

| Distance to receiver (ly) | Flux in 100 yr (ph m⁻² s⁻¹) | Flux in 10⁶ yr (ph m⁻² s⁻¹) |
|---|---|---|
| 0.01 | 5.1 x 10¹¹ | 5.1 x 10⁷ |
| 0.1 | 5.1 x 10⁹ | 5.1 x 10⁵ |
| 1 | 5.1 x 10⁷ | 5,100 |
| 10 | 5.1 x 10⁵ | 51 |
| 100 | 5,100 | 0.51 |

The communication bit rate cannot be significantly higher than the photon rate, as the presence or absence of a photon can convey no more than a 1 or 0. We ignore phase information, which hardly changes this constraint. Thus, the photon rates given by equations 2 and 3 and listed in Table 1 are essentially equal to the bit rates of information for meter-class probes. These benchmark bit rates of ~50 Mbps are modest, less than common internet speeds, and thus merit examination.

Actual probes could have properties to enhance the bit rate over the benchmark values. For example, actual probes could have 10x or 100x greater initial energy. Equation 3 shows that such enhancements in initial energy, $E_i$, would augment the bit rate proportionally from those listed in Table 1. Of course, such enhanced energy payloads come at the price of increased fuel, mass, and cost, and decreased maneuverability.

Actual probes could have beam launchers 10× larger, i.e., D =10 meters in diameter. Equation 3 shows that the resulting bit rate increases as $D^4$, achieving 10,000 × higher bit rate. As a last example, the wavelength emitted by the launcher could be several orders of magnitude larger or smaller. Equation 2 shows that the resulting bit rate increases inversely with wavelength, thereby favoring smaller wavelengths.

The examples above exemplify actual interstellar probes having properties different from the benchmark values in Table 1 by several orders of magnitude, yielding similarly different bit rates for communication. Still, these bit rates for distances over 1 ly are no more than the highest internet speeds, assuming a probe transmitting lifetime of only a century. For probe lifetimes of $10^3$ to $10^6$ years, the resulting bit rates are proportionally slower, typically below 1 Gbps. These bit rates are idealized, optimal achievements, limited only by the physics of energy and optical diffraction. Engineering realities may yield lower bit rates than those implied by Equation 3 and Table 1.

Probe-to-probe bit rates over distances of a light year are severely constrained by energy budget and the diffraction of light. Of course, diffraction is equivalent to the Heisenberg uncertainty principle, causing uncertainty in the momentum of the photon. Thus diffraction in Equation 3 is incontrovertible, fundamental physics as we know it.

In addition, interstellar communication probes pose engineering challenges in constructing the probe, its propulsion, its precision optics, its energy generation, and its station-keeping. Advanced technologies may meet these engineering challenges. But the physics that constrains bit rates remains for communication over distances of a light year, never mind greater distances. The relevant distance scale is the separation of stars - typically 3 to 10 light years apart.

The bit rates derived here from physics are so modest that they limit the amount of information a probe might send to its recipients about Earth and its inhabitants. Bit rates of no more than 1 Gbps are what we humans want with a "5G" internet. Such bit rates are hardly enough to transmit all the key information about the entire Earth to the rest of the Galaxy.

Any probe that happens to be near a host star might supplement its energy by photovoltaic cells or by excavating material from rocky asteroids or planets to feed a nuclear reactor. Regarding the energy budget, the probe might be in an elliptical orbit around the host star and a habitable planet. When close to the star it could harvest photons with photovoltaic cells and store the energy in batteries. When close to the planet, the probe could conduct surveillance.

Such supplementary energy, collected over the lifetime of the probe, must be significant relative to payload energy. However, as we describe below, some (or most) probes may be nowhere near a star system, unable to augment their initial endowment of energy.

## 2.2 Stellar Gravitational Lensing (SGL)

In principle, the bit rate might be solved if a local ET probe transmits its information through a solar gravitational lens (SGL) [12, 13, 14]. In accordance with Einstein's Theory of General Relativity, a beam of light originating from a spot on a focal line that begins 550 AU from the host star (for a Solar-type star) and traverses just outside the perimeter of the star would be bent into a sharp focus at some distant spot. The theoretically achieved gain would be on the order of $10^9$ and the increase in bit rate would be on the order of $10^6$. The local surveilling probe might transmit data to a sister probe located at >550 AU, which in turn, would use the SGL effect to transmit data to a distant star. However, in practice, gravitational lensing to improve the gain might be difficult to implement for the following reasons:

- The SGL focal region has a size of only a few meters for optimal "gain" [8, 9]. Displacements of only a few kilometers from the focal point diminishes the gain by many orders of magnitude.

- The Sun, as a lens, is not stationary, but instead moves in a curly-cue pattern by ~1 million kilometers during several years because the planets gravitationally yank on the Sun. A probe intending to reside at the focus of the Solar gravitational lens must therefore move billions of kilometers over thousands of years using precious propulsion fuel to "shadow" the Sun's odd motion. That is, station keeping is challenging and energy intensive as both the distant star and the local one (e.g. the Sun) move millions of kilometers due to their planets yanking gravitationally on them. Binary stars cause yet more problems [15,16].

- Aiming the emitter at the receiver within ~1 meter, over distances of many light years, is technically challenging.

- The solar corona adds noise to signals that are grazing just outside the surface of the Sun, especially radio, optical, and near-IR wavelengths at which the Sun is bright.

## 3   REPEATER STATIONS

Another solution stems from considering repeating nodes placed at small intervals between stars, analogous to repeaters in telephone cell towers. Table 1 shows that benchmark meter-class probes separated by 0.01 light year can send 50 million photons per meter$^2$ per second to each other, with a duration of a million years. Figure 2 shows a sketch of one architecture of the probes between the Solar System and Alpha Centauri. The train of probes emit diffraction-limited beams carrying bits in the form of photons. The closer the spacing of the probes the higher the bit rate of information, by limiting light losses as the square of the distance.

Using the example of repeater nodes at 0.01 light years separation between Earth and Alpha Centauri, approximately 420 nodes would be required. Such nodes would require a steerable transmitter and receiver, a tracking system (small telescope), onboard CPU, propulsion system for course adjustment, and a reliable energy supply.

Each repeater node might traverse the Galaxy going from one star to the next, course correcting as it approaches each star and then sling shotting towards the next star. An alternative is a string of pearls permanently placed between two stars, roughly 0.01 ly apart, and adjusting positions slightly to account for the steady velocity vector of the two stars as they move through their galactic neighborhood. The probes may need a modest propulsion system that allows them to remain indefinitely in a specific location relative to each star. For example, the first probe would be placed at 0.01 ly from one star to the next, the second at 0.02 ly, the third at 0.03 ly, and so forth, with modest station keeping. In the event that a communication node becomes inoperable or depletes its fuel, the nodes on either side of it might skip it and communicate directly with each other. Assuming even spacing, this would result in an immediate 4-fold decrease in bit rate, however, entire strings of nodes might re-space themselves reducing this gap and increasing the bit rate towards its original design.

Under the terms described in this paper, the local ET probe would have already surveilled Earth's electromagnetic leakage, and taken from it vast amounts of data. Because the probe already is in the possession of it, we would not be able to trade our information for the proverbial *Encyclopedia Galactica*. However, humankind may have an ability to negotiate for ET's vast reservoir of information in return for services the probe might require it to render. The probe may require human assistance in maintaining the local galactic Internet by building probes and especially replacement communication nodes for those that have become dysfunctional. Indeed, we might be called upon to build out whole sections of the galactic Internet, for example, by building and launching an entire suite of communication nodes between Sol and a nearby star with which no direct link currently exists.

The construction of communications nodes may be well within humankind's current manufacturing capabilities. These nodes might even be mass-produced and even be no more expensive to produce than a luxury car. They might be routinely launched from low gravity surfaces like the Moon, Mars or asteroids. Although they will need to be accelerated to the solar system escape velocity (42 km/sec if launched from the Moon or 19 km/sec if launched from 5 AU), no large further velocity would be required and might be counterproductive as faster speeds entail more danger from collisions with micrometeor-ites or dust particles. Given that Voyager is currently only ~22% of the way to the first proposed station of 420 (were it headed toward Alpha Centauri, which it is not), this would entail planning and execution over deep time. Alternatively, nodes might be launched at an appreciable percentage of the speed of light, but this would require much more energy to launch and decelerate into station keeping. Fully robotic AI probes, as opposed to repeater nodes, may be beyond our current ability to manufacture because they are assumed to possess very advanced computers, although ET might give us design and manufacturing instructions.

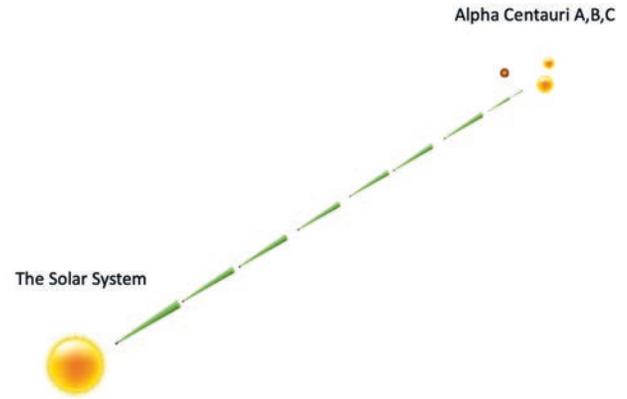

**Fig.2**  Sketch of a series of probes serving as repeater stations between the Solar System and Alpha Centauri. The probes emit diffraction-limited light beams carrying high bandwidth information, similar to fiber optic cables or satellite-based internet. Close probe spacing reduces bit-rate loss proportional to the square of distance between them (see Equation 3).

## 4   RECRUITING EARTH TO ACT AS A NODE

A local ET probe might seek to recruit humans to transmit directly to other civilizations, since we have much larger telescopes, and have access to larger quantities of energy than are likely to be immediately available to the probe. We, in effect, would become its node. It might instruct us on how to communicate by means of giant terrestrial lasers and 10-meter-class optical telescopes or kilometer-class (FAST, MeerKat, Square Kilometer Array) radio telescopes to the nearest civilizations. The probe might persuade us to transmit the intelligence it had gathered on Earth but not yet transmitted to its progenitor civilization in a format that the progenitor might understand, but that might be unintelligible to us.

We ourselves could become a repeater node between remote civilizations. We might be instructed to forward a message from one star at, say, 500 ly, to another star in the opposite direction at 500 ly (total 1,000 ly), even though the messages might be in an incomprehensible (to us) format. In other words, we would be acting merely as a relay station between civilizations without having any idea of what the content of the message might be. With the assistance of the local probe, we, in turn, might be linked to civilizations with whom we might intelligibly communicate, though perhaps also through a network of nodes rather than directly. Our messages would go back and forth though the same Internet being relayed by civilizations with whom we might have nothing to speak about, but who nevertheless act as nodes for us, just as we function as a node for them.

### 4.1  What Would Be ET's Preferred Engineering Solution?

The short answer to the question posed in this header is that we cannot know for sure. Every solution requires tradeoffs. We believe that the SGL solution, though theoretically interesting, may be too difficult to achieve in practice. Small probes separated by the distance of even the nearest stars will have a bit rate that is incompatible with the vast amount of information the probes would presumably seek to transmit, especially after it had detected Earth's artificial electromagnetic radiation. Therefore, ET might prefer large probes transmitting in such high frequencies as UV, using large transmitters and receivers rather than the 1 m class transmitters and receivers we have used as a benchmark, perhaps on the scale of a medium size asteroid, and perhaps already catalogued but mischaracterized by astronomers as an asteroid. In that event, such a large probe in our Solar System could be in direct communication with sister probes around such nearby stars as Alpha Centauri. Alternatively, the solutions set forth in Sections 2.3 and 2.4 above may be ET's least difficult, and therefore most probable solution. Small probes would have been easier for ET to manufacture, launch, and maintain within our Solar System over deep time. But in that case, the efficient transmission of information back to its progenitor civilization may require the recruitment of humankind.

### 5  COMMUNICATIONS BETWEEN A LOCAL ET PROBE AND EARTH

Each of the solutions proffered in Sections 2.3 and 2.4 above would require that the local ET probe open a channel of communications with Earth. Having surveilled and recorded vast amounts of information about Earth's single technological species, *homo sapiens*, the ET might be able to initiate communications with Earth in English, Chinese, or some other terrestrial language that it had learned. By means of this communication it might entice us to participate in the maintenance of interstellar communications. Such first contact might take the form of a laser, such as given as an example in Table 1, which would produce a light cone diameter of 95 km on the face of the Earth if transmitted from a distance of Mercury. If transmitted from the distance of the Moon, the light cone would cover a mere 245 m. From the distance of Mercury, its apparent magnitude would be -11 while from the distance of the Moon, it would be -23. In either case, it would be not just a naked eye object, but by far the brightest point source in the night sky.

The beam should be almost instantaneously recognized as extraterrestrial by other measures as well as its luminosity. Unlike an airplane, satellite or planetary or stellar object, the beam could be monochromatic. Unlike an airplane, satellite, or planetary or stellar object it will only be visible within a small light cone rather than visible to any observer above its horizon.

The light cone may move over the surface of the Earth if the probe's beam is fixed relative to itself, illuminating a circumterrestrial ribbon like a solar eclipse trail. Alternatively, the beam might track a single location on the Earth, perhaps an astronomical observatory or landmark, such as the Greenwich Observatory, whose location is known to the probe and fundamental to Earth's coordinate system. When tracing the beam backwards to its source, the probe will be seen to move in a fashion that is completely different from a satellite or airplane. If it is en route from the Sun, in a highly eccentric elliptical orbit that also includes the Earth, then, like a comet it will not move much from moment to moment. Once identified (almost immediately) as an alien object, proper instrumentation, including CCD and CMOS sensors and radio receivers, can be placed within the light cone to download its message.

To properly surveil Earth's leakage of electromagnetic radiation, the local ET probe will have to have been equipped with a radio receiver. Therefore, it is possible that ET will transmit its message to Earth in the radio spectrum, possibly taking over bands that are normally reserved for AM, FM or TV broadcasts. Any radio beam would have a very wide footprint, depending on the probe's distance from Earth and the size of ET's transmitter, it may even cover the entirety of Earth's facing hemisphere.

### 6  CONCLUSIONS

We have demonstrated that probes within the Milky Way Galaxy designed to perform surveillance and to communicate their acquired information over interstellar distances face constraints from fundamental principles of physics. The maximum possible energy budget for generating electromagnetic communication and the diffraction of the transmitted beam impose physics-based limits on the rate of information that can be transferred from one probe to another.

We derived an expression, Eq. 3, that gives the optimal information bit rate between two probes of a given size, separation, lifetime, and wavelength. The highest information bit rate possible over distances of a light year for meter size probes having life spans of 100 yr to $10^6$ yr, is roughly a gigabit per second, depending on life span, as given in Table 1. We consider alternative probe architectures, including stringing over 100 "repeater" probes between stars to constitute an interstellar conduit of communication, functionally analogous to fiber optics.

A string of interstellar repeater nodes (Fig. 2) offers the key advantage of tightening the communication beams that otherwise waste energy and lower bit rate as the square of transmission distance. The bit rate increases inversely as the square of the node spacing, for a given energy budget and transmitter size (Eq.3). The desirable densification of nodes for interstellar communication raises two issues immediately. There is an obvious trade-off between the node density (nodes per light year) and the cost of manufacturing and maintaining those nodes. There is also a trade-off between node density and the size of the optics. Bigger optics allow for fewer nodes to achieve the same bit rate. Without knowing the cost of such nodes, and an associated funding budget and business model, it is difficult to guess the resulting optimized density and size of the nodes.

The engineering of a linear string of nodes has to plan for a single-point failure. Losing functionality of just one node might cause the entire communication linkage to fail. We suggest solutions in Section 2.3 above. However, it is also possible that the topology of the interstellar network could improve fault tolerance against single-point failures, and yet satisfy the cost-benefit trade-offs for one-dimensional star-to-star communication [17]. For example, the same message that a local probe intends for its progenitor civilization many light years away might be sent through very many routes simultaneously; in one pathway, the first leg would go through Alpha Centauri, while in others the first legs would be routed through different nearby stars.

Humanity is unprepared for contact with extraterrestrial technology in general, and most particularly from contact with an ET probe in Earth's immediate vicinity. Most immediately and urgently, small and portable telescopes with efficient sensors (e.g. sCMOS or CCD) and radio telescopes with back-end recording devices must be built and prepositioned on Earth so that they can be very quickly brought to bear on an alien radio or laser beams emanating from within our own Solar System. An effective laser communication system should be designed and built for the purpose of responding to the ET probe. More urgent still, policy decisions should be taken in advance of first contact that encompass as many contingencies as possible, and these should be enshrined in international treaties. Thrashing out our response to each contingency might take years, but better we spend those years in advance rather than with dumbfoundness in the immediate aftermath of a local ET probe detection.

### Acknowledgements

We thank Nate Tellis, Michael Garrett, Michael Hippke, Beatriz Villarroel, Martin Ward, and Andrew Siemion for helpful suggestions.